\documentclass[%
 reprint,
 amsmath,amssymb,
 aps,
 pre,
]{revtex4-1}

\usepackage{graphicx}
\usepackage{dcolumn}
\usepackage{bm}
\usepackage{url}

\usepackage{amsmath}
\usepackage{appendix}
\usepackage[bf]{subfigure}
\usepackage[utf8]{inputenc}
\usepackage{tikz}

\newcommand{\mean}[1]{\langle #1 \rangle}

\newcommand{\inst}[1]{$^#1$}
\newcommand{\pavoid}{p_{a}}

\newcommand{\khomix}{k_H}
\newcommand{\Rhomix}{R_H}
\newcommand{\ttransient}{t_{tr}}
\newcommand{\taverage}{t_{av}}
\newcommand{\nexec}{n_{ex}}

\newcommand{\homix}{HM}  
\newcommand{\rw}{RW}  
\newcommand{\lr}{LR}  

\begin{document}

\preprint{APS/123-QED}

\title{Epidemic spreading in populations of mobile agents with adaptive behavioral response}

\author{Paulo Cesar Ventura\inst{1}}
\author{Alberto Aleta\inst{5}}
\author{Francisco A. Rodrigues\inst{2}}
\author{Yamir Moreno\inst{{3,4,5}}}

\affiliation{
  \inst{1} Instituto de F\'{i}sica de  S\~{a}o Carlos, Universidade de S\~{a}o Paulo, S\~{a}o Carlos, SP, Brazil.\\      
   \inst{2} Instituto de Ci\^{e}ncias Matem\'{a}ticas e de Computa\c{c}\~{a}o, Universidade de S\~{a}o Paulo, S\~{a}o Carlos, SP, Brazil.\\
    \inst{3}Institute for Biocomputation and Physics of Complex Systems, University of Zaragoza, Zaragoza 50018, Spain\\
  \inst{4}Department of Theoretical Physics, Faculty of Sciences, University of Zaragoza, 50009 Zaragoza, Spain\\
  \inst{5} ISI Foundation, Turin, Italy
}

\begin{abstract}

Despite the advanced stage of epidemic modeling, there is a major demand for methods to incorporate behavioral responses to the spread of a disease such as social distancing and adoption of prevention methods. Mobility plays an important role on epidemic dynamics and is also affected by behavioral changes, but there are many situations in which real mobility data is incomplete or inaccessible. We present a model for epidemic spreading in temporal networks of mobile agents that incorporates local behavioral responses. Susceptible agents are allowed to move towards the opposite direction of infected agents in their neighborhood. We show that this mechanism considerably decreases the stationary prevalence when the spatial density of agents is low. However, for higher densities, the mechanism causes an abrupt phase transition, where a new bistable phase appears. We develop a semi-analytic approach for the case when the mobility is fast compared to the disease dynamics, and use it to argue that the bistability is caused by the emergence of spatial clusters of susceptible agents. Finally, we characterize the temporal networks formed in the fast mobility regime, showing how the degree distributions and other metrics are affected by the behavioral mechanism. Our work incorporates results previously known from adaptive networks into population of mobile agents, which can be further developed to be used in mobility-driven models.

\end{abstract}


\maketitle

\section{Introduction}

 

Epidemics are of great concern to humankind. While it is possible to construct realistic epidemic models with heavy use of data, the effect of human behavior, from individual to collective level, is crucial to the dynamics but difficult to be incorporated into models. It is known that mobility patterns, as an important part of human behavior, strongly influence the spread of a disease, and mobility data from real world is often incorporated to epidemic forecasting.

Despite the increasing availability of data from human mobility \cite{eagle2006reality,cattuto2010dynamics}, there are several situations in which such data is not available. Therefore, the use of 
synthetic mobility models 
to feed epidemic modeling \cite{gonzalez2004scaling,frasca2006dynamical,buscarino2008disease,buscarino2010effects,zhou2009epidemic,zhou2012epidemic_mobility,buscarino2014local,ichinose2018reduced,Huang_2016,peng2019sis,feng2018epidemic,fofana2017mechanistic} is a promising, yet understudied topic. Note that in this context, mobility refers to short-range displacement of individuals - such as walking - and not to the transfer of individuals from one place to another.
Frasca \cite{frasca2006dynamical} \emph{et al.} and Buscarino \cite{buscarino2008disease} \emph{et al.} proposed an SIR model in a population of random walking agents that can also perform long-range jumps. They study the relationship between final outbreak size and some average topological properties of the underlying time-aggregated network of contacts, such as degrees, shortest path lengths and clustering. Interestingly, the long-range jumps cause a small-world effect similar to the Watts-Strogatz model \cite{watts2004small}, which seems to explain the increase in the outbreak size with the jumping probability. Their work settled the base for epidemic modeling in populations of random walkers. 

One of the major difficulties of working with epidemic spreading with random walking agents is that the spatial component imposes a temporal correlation to the contacts, which strongly influences the epidemic spreading. For very slow mobility, the population can be regarded as a static network, whereas for fast mobility the correlations are broken and the system is essentially homogeneously mixed. For this reason, despite some insights from reaction-diffusion processes \cite{wang2010travelling}, the mid-term between these regimes essentially relies on computational Monte Carlo simulations. More recently, other works have considered spatial heterogeneity and separate communities \cite{zhou2009epidemic,zhou2012epidemic_mobility,buscarino2014local}, heterogeneous interaction radii \cite{Huang_2016,feng2018epidemic, peng2019sis} and different agent's velocities \cite{buscarino2010effects,ichinose2018reduced}.

Despite the inherent difficulties, mobile agent models for epidemic spreading can be very insightful, as some models can reproduce features of real world human and animal mobility. Starnini and others \cite{starnini2013modeling} modified the basic random walk to consider heterogeneous activity time and attractiveness of the individuals, successfully reproducing behaviors of some data sets of the SocioPatterns collaboration \cite{cattuto2010dynamics}. Stehl{\'e} and others \cite{stehle2011simulation} performed SEIR simulations using interactions between attendees of a conference, and showed that a correct estimation of the epidemic dynamics must account for the  duration of contacts, a feature that is in parts reproduced by Starnini's model. Another important example is the Levy walk, a modification of the simple random walk that considers heavy-tailed step size distributions. Despite some recent evidences claim for the use of more complex models \cite{pyke2015understanding,reynolds2015liberating}, Levy walks have been widely used to describe animal and human mobility patterns \cite{ramos2004levy,rhee2011levy}.

\begin{figure*}[ht]
    \centering
    \includegraphics[width=0.7\textwidth]{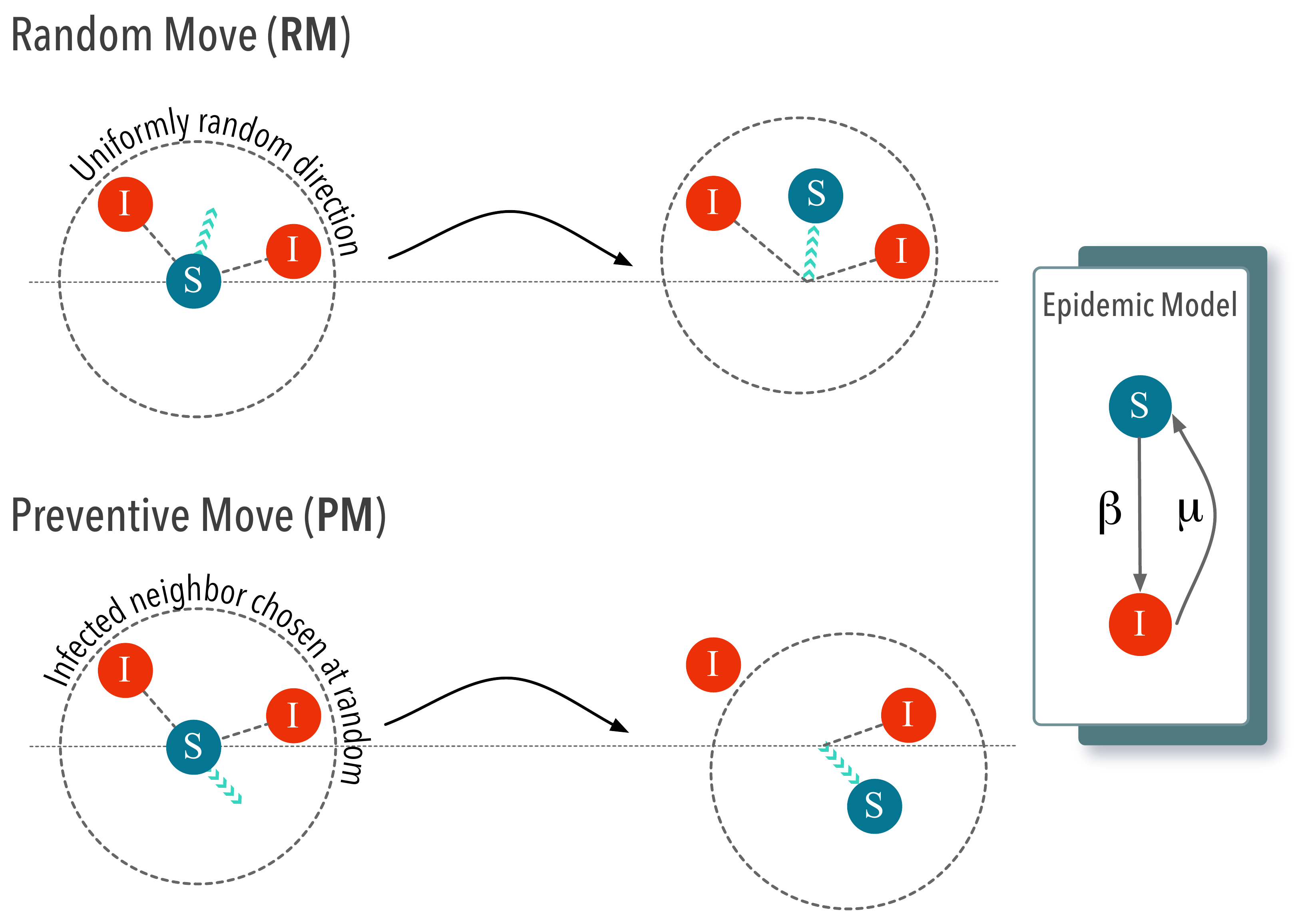}
    \caption{Schematic representation of the possible mobility steps and the epidemic model. A susceptible (S) agent can either perform a random move (RM) to a direction chosen at random, or a preventive move (PM), in which it chooses one of its infected neighbors (if more than one is present) and moves away from it. }
    \label{fig:model_scheme}
\end{figure*}

In spite of the importance of mobility models for disease dynamics, little has been done to include behavioral changes into such models. Human responses crucially influence the spread of a disease, and there is an intense research devoted to account for such effect into epidemic models, which include responses based on risk perception\cite{risk_perception_multiplex,poletti2012risk,wu_2012}, spreading awareness and individual prevention methods\cite{arenas_2014,Funk6872,da2019epidemic}, vaccination games \cite{liu2012impact,cardillo2013evolutionary} and others\cite{wang2015coupled}. 

One particularly popular approach is to consider adaptive contacts in networks, which mimic both the social distancing and isolation of infected individuals. Gross and others \cite{gross2006epidemic,gross2008robust} proposed a simple model in which susceptible individuals can randomly rewire their links that point to infectious nodes, redirecting them to other susceptibles. This generates a dynamic network that evolves simultaneously to the disease spreading, with susceptibles forming highly connected clusters between themselves to exclude infecteds. This behavior produces rich dynamical phenomena such as bistability, oscillations and hysteresis. The model later received refined analytical treatments \cite{marceau2010adaptive,guerra2010annealed} and motivated a series of subsequent studies \cite{shaw2008fluctuating,risau2009contact,zhou2012epidemic,wieland2012structure}. In particular, Zhou and others \cite{zhou2012epidemic} proposed a variant of the adaptive rewiring model that considers network growth and isolation avoidance, showing that the combination of these two factors can produce multiple epidemic bursts in an SIS model before the disease is eradicated. 


In the present work, we propose to merge the richness of adaptive behavioral responses with the modeling potential of mobile agents. We propose a mechanism through which susceptible agents can avoid contact with infected agents by performing preventive moves. As the model itself considers a very simple adaptive mechanism, in which susceptibles are fully and instantaneously aware of the state of their neighbors, the aim of this work is not to provide results that are directly applicable to the real world. Instead, our goal is to propose the merge between mobility models and adaptive responses, study its basic dynamical properties and motivate future developments. 

We also use a simple semi-analytical approach for the case in which the disease dynamics is slow compared to the agents' mobility. While it still relies on computational simulations, it allows to easily extract, among other results, the stationary prevalences and the phase diagrams, including the transcritical and saddle-node bifurcations that our model presents. Ultimately, it also provides an interpolated functional form for the reduction of contacts due to the adaptive mechanism. This can be applied into the local/regional dynamics of more complex models, such as those with metapopulations \cite{lloyd1996spatial,colizza2008epidemic,meloni2011modeling,aleta2017human,grenfell1997meta,iannelli2017effective}, to include behavioral responses to epidemic spreading.


\section{The model}

\subsection{Epidemic model}

The epidemic model employed in this work is the reactive \emph{Susceptible-Infected-Susceptible} (SIS) with discrete time evolution. Each agent can either be susceptible (S) to the disease or infected (I). At each time step, an infected agent that interacts with a susceptible one can transmit the disease with probability $\beta$. In addition, an infected agent can also be healed with probability $\mu$. Infection and healing events are only applied in the next time step, so the order of agent visits does not matter, and reinfection after healing in a single time step is not allowed.

\subsubsection{Basic mechanism of motion and interaction}

For the baseline population dynamics, we use the simple random walk with hard interaction circles, as in references \cite{frasca2006dynamical,buscarino2008disease,buscarino2010effects}, yet with no long-range jumps. A population of $N$ agents is initially distributed at random in a square space of length $L$ with periodic boundary conditions. At each time step, each agent can perform a \emph{random move} (RM) of fixed length $v$ and uniformly random direction $\theta$. Thus, the horizontal ($x$) and vertical ($y$) coordinates of the position of the agent at time $t + 1$ after a random move are given by:

\begin{equation}
    \begin{cases}
    x(t+1) = x(t) + v\cos(\theta)  \\
    y(t+1) = y(t) + v\sin(\theta)
    \end{cases}
\end{equation}


Each agent has an interaction radius $r$, meaning that if two agents have a spatial distance smaller than $r$, they interact reciprocally. With this interaction scheme, one can construct a snapshot network at each time step of the model.

To avoid any dependency on the implementation, both epidemic and positional state changes are calculated at each time step, but they are only applied after all changes were calculated.

\subsubsection{Local reaction mechanism}

\begin{figure}
    \centering
    \includegraphics[width=0.95\columnwidth]{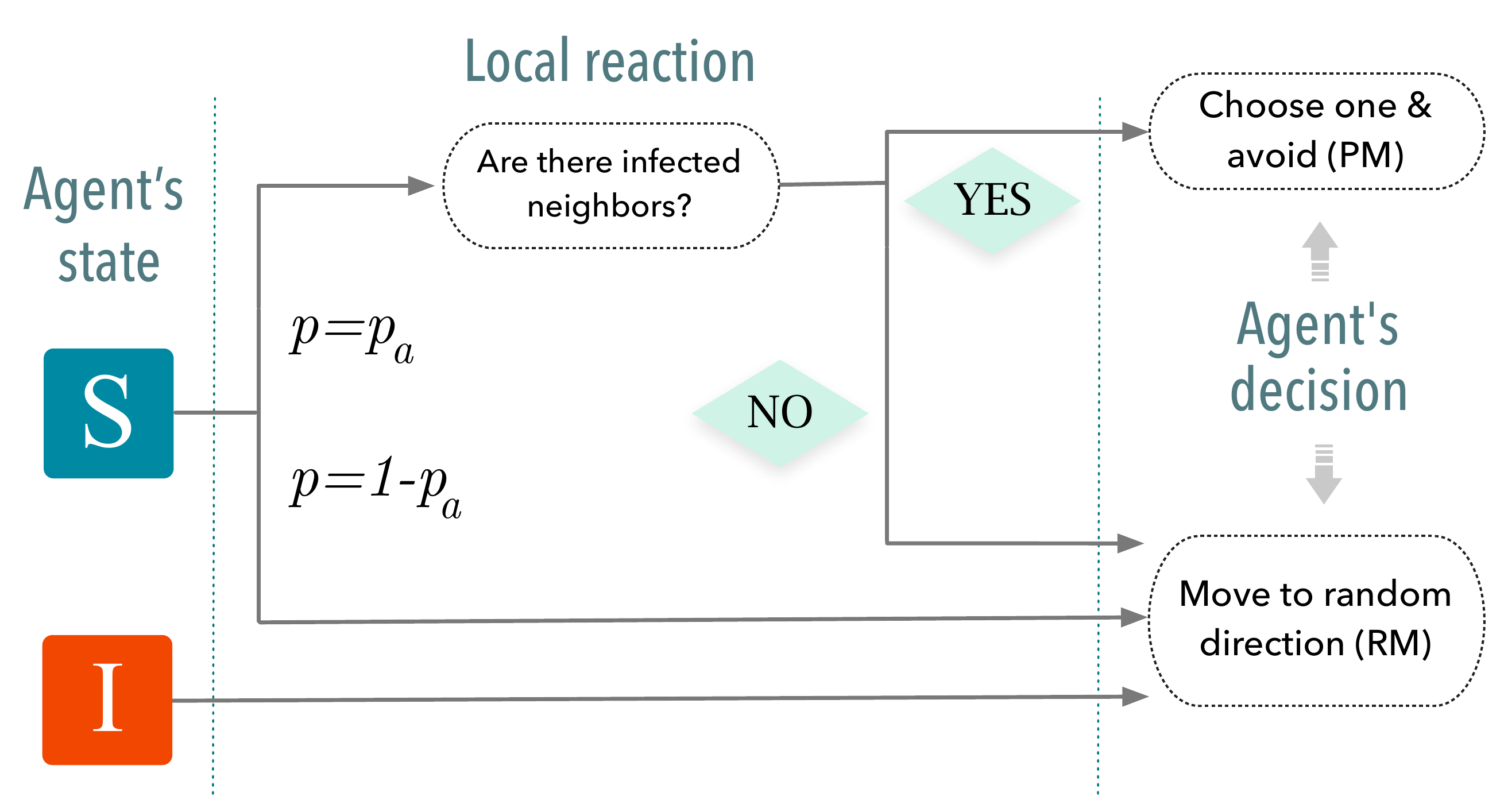}
    \caption{Scheme of the decision algorithm for the motion of each agent, at each time step. With the local reaction mechanism, susceptible agents may decide to move away from an infected neighbor, if there is one, with probability $\pavoid$.}
    \label{fig:motion_diagram}
\end{figure}

We incorporate an adaptive reaction to the local presence of infected individuals. At each time step, each susceptible (S) individual chooses, with probability $\pavoid$, to avoid its interaction with an infected neighbor by moving away from it, using the following algorithm: (i) if there is exactly one infected neighbor, the susceptible agent moves in its opposite direction with step size $v$, which we call a \emph{preventive move} (PM). Here opposite direction means that the displacement vector makes an angle of $\pi$ with the relative position vector that goes from the S to the I agent (see figure \ref{fig:model_scheme}. (ii) If there are two or more infected neighbors, the susceptible chooses one of them at random and promotes a preventive move away from it. (iii) If there are no infected neighbors, it simply promotes a random move (RM) to a uniformly random direction, as already described. Also, with probability $1 - \pavoid$, the susceptible promotes a random move regardless of its infected neighborhood. Figure \ref{fig:model_scheme} illustrates the PM and RM actions, as well as the SIS epidemic model, while figure \ref{fig:motion_diagram} shows the agents' algorithm for choosing between PM and RM.

Notice that, also as a simplification, the ``awareness radius'' of the agents is the same as the interaction radius $r$. Another simplification is that every infected individual is immediately perceived as such, which can be interpreted, for example, as if the disease symptoms are clear and always display immediately after infection.




\begin{table}
\caption{\label{tab:table1} List of symbols}
\begin{ruledtabular}
\begin{tabular}{lcr}
Symbol&Meaning\\
\hline
$N$       & Number of agents                              \\
$N_I$     & Number of infected agents\\
$L$       & Size of the square space                      \\
$r$       & Interaction radius of the agents              \\
$v$       & Size of spatial step                          \\
$\pavoid$ & Probability of avoiding infected neighbors \\
$\ttransient$ & Simulation transient period                   \\
$\taverage$ & Simulation stationary period (data collection) \\  
$\nexec$    & Number of independent executions    \vspace{0.5cm} \\
$\beta$     & Infection probability                     \\
$\mu$       & Healing probability                       \\
$\lambda$   & $= \beta / \mu \quad $ Infection-to-healing ratio \\
$\khomix$   & $= (N / L^2)\pi r^2 \quad$ Homogeneous degree  \\
$\Rhomix$   & $= \lambda \; \khomix \quad$ Homogeneous reproduction number \\
$\rho_{I}$  & $= N_I / N \quad$ Disease prevalence \\
$\rho_{I}^*$  & Steady state disease prevalence \\
\vspace{0.5cm} \\
\homix{}    & Homogeneously mixed population                  \\
\rw{}       & Random walker agents                                  \\
\lr{}       & Random walk + local reaction agents                                 \\
PM          & Preventive move               \\
RM          & Random move       \\
\end{tabular}
\end{ruledtabular}
\end{table}

We perform Monte Carlo simulations of the SIS model under two different motion schemes for the mobile agents: (i) simple random walk (\rw), for which no reaction mechanism is considered (i.e., $\pavoid = 0$) and (ii) random walk with local reactions (\lr), for which we use $\pavoid = 1$, except where explicitly mentioned. 
We also compare our model with a homogeneously mixed population (\homix) with an average number of contacts per unit time given by: 

\begin{equation}
    \label{eq:k_homix}
    \khomix = \frac{N}{L^2} \cdot \pi r^2
\end{equation}

Which is the expected average degree of a population of agents uniformly distributed in space. We call $\khomix$ the \emph{homogeneous degree}. We can also define the \emph{homogeneous reproduction number} (i.e., the epidemic reproduction number if the nodes were homogeneously mixed with $\khomix$ contacts per time step in average) as: 

\begin{equation}
    \Rhomix = \lambda \khomix
\end{equation}
\noindent
where $\lambda = \beta / \mu$ is the infection-to-healing ratio of probabilities. 


\section{Basic results}
\label{sec:basic_results}

We start the analysis by studying the stationary state prevalence $\rho_I^*$ of the system under a fixed initial infected fraction $\rho_I(0)$. We perform simulations under four different regimes with respect to the density of agents and the relative time scale between epidemic and motion dynamics. Henceforth we call slow epidemics to those in which $\mu \ll v$, while for epidemics with similar time scale between the spreading and motion dynamics we have $\mu \sim v$.

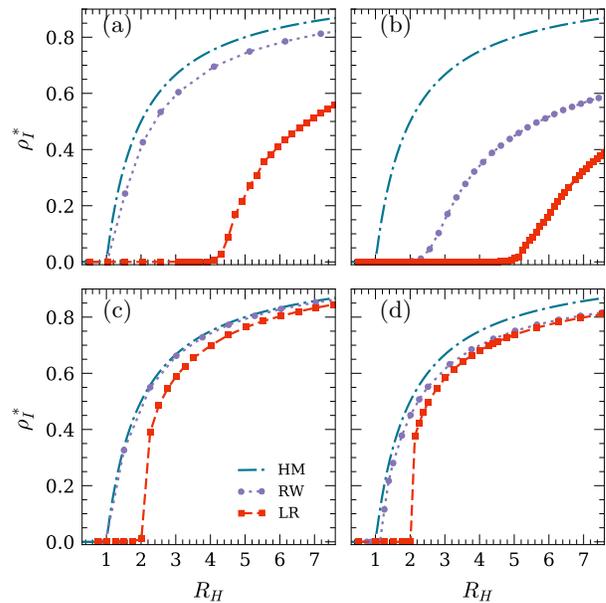
\begin{figure}
    \begin{tikzpicture}
        \node[above left] (img) at (0,0) {\includegraphics[width=0.95\columnwidth]{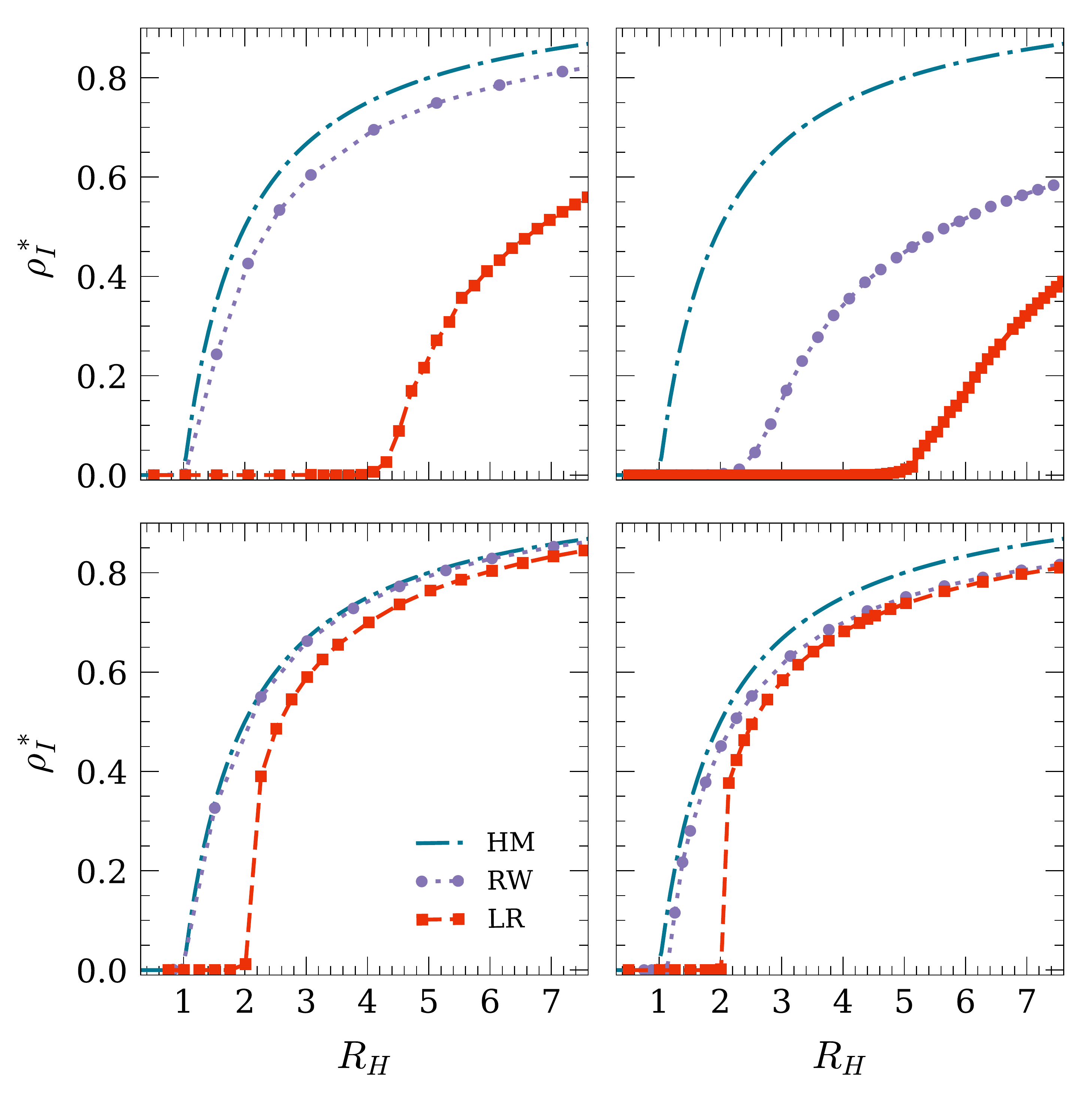}};
        
        \node[text=black] at (-6.8, 8.0) {(a)};
        \node[text=black] at (-3.1, 8.0) {(b)};
        \node[text=black] at (-6.8, 4.2) {(c)};
        \node[text=black] at (-3.1, 4.2) {(d)};
        
    \end{tikzpicture}
    \caption{Stationary prevalence as a function of the homogeneous reproduction number $\Rhomix$, for different models of population (HM, RW and LR), under four different regimes: (a) low density, slow epidemics; (b) low density, similar time scale; (c) high density, slow epidemics; (d) high density, similar time scale. High (low) density is obtained by setting the $L = 10 \, (35)$, for which the homogeneous degree is $\khomix = 12.57 \, (1.03)$. The slow epidemics is performed with $\mu = 0.005$, while the similar time scale regime uses $\mu = 0.1$. Other parameters are: $N = 400$, $r = 1.0$, $v = 0.3$. Different values of $\Rhomix$ are obtained by varying the infection probability $\beta$. Each curve is averaged over $\nexec = 50$ or $500$ executions for the slow and similar time scale models, respectively.}
    \label{fig:infective_curves}
\end{figure}

In figure \ref{fig:infective_curves}, we show the stationary prevalence as a function of the homogeneous reproduction number $\Rhomix$, for $\rho_I(0) = 0.30$. In each execution, the model is first simulated for a transient period of $\ttransient$ time steps, and then for $\taverage$ additional steps during which the data is collected and averaged. Each point is also an average of a number $\nexec$ of such independent executions. 

\begin{figure}[h]
    \centering
    \includegraphics[width=0.95\columnwidth]{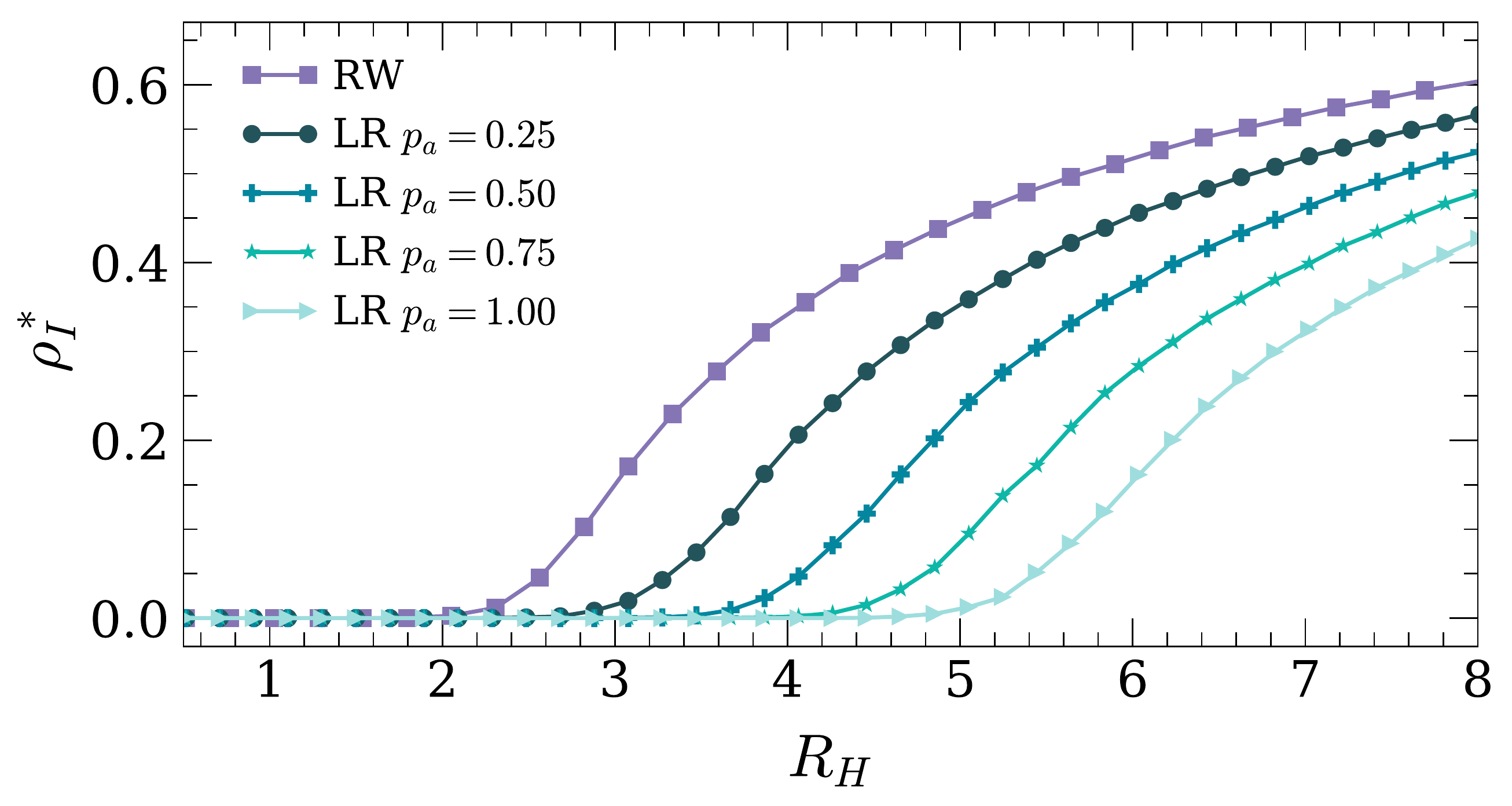}
    \caption{Stationary prevalence $\rho_I^*$ as a function of the homogeneous reproduction number $\Rhomix$ in the low density and similar time scale regime (as in figure \ref{fig:infective_curves}.b), for different values of the avoidance probability $\pavoid{}$. Parameters are: $\mu = 0.1$, $r = 1$, $v = 0.3$, $L = 35$, $N = 400$ (thus $\khomix = 1.03$). Each curve is an average over $\nexec = 500$ executions.}
    \label{fig:vary_pa}
\end{figure}

First analyzing the low density regime (panels (a) and (b)), we notice that the local reaction mechanism (\lr) considerably raises the epidemic threshold, both with respect to homogeneous mixed (\homix) and random walking (\rw) populations. 
In the high density regime (panels (c) and (d)), we notice an abrupt phase transition from healthy to endemic stationary states with the \lr{} population, which is not observed within the other models. Such phenomenon was already reported for SIS-like models on adaptive networks \cite{gross2006epidemic,shaw2008fluctuating,risau2009contact}, and is actually a fingerprint of another important feature: a bistable phase caused by a saddle-node bifurcation.

Yet from figure \ref{fig:infective_curves}, we notice that our \lr{} model deviates from homogeneous mixing even in the condition of slow epidemics (panels (a) and (c)), while the simple random walk (RW) can be reasonably described by \homix{} in this regime. 

We evaluate the effect of the probability of avoiding infectious contacts $\pavoid{}$ in figure \ref{fig:vary_pa}, which shows the prevalence curves at the low density and similar time scale regime for different values of $\pavoid{}$. For intermediate values of the parameter, the curves monotonically fill the range between full avoiding mechanism (\lr{} $\pavoid{} = 1$) and simple random walk (\rw{}, equivalent to \lr{} with $\pavoid{} = 0$).

\begin{figure}
    \centering
    \includegraphics[width=0.95\columnwidth]{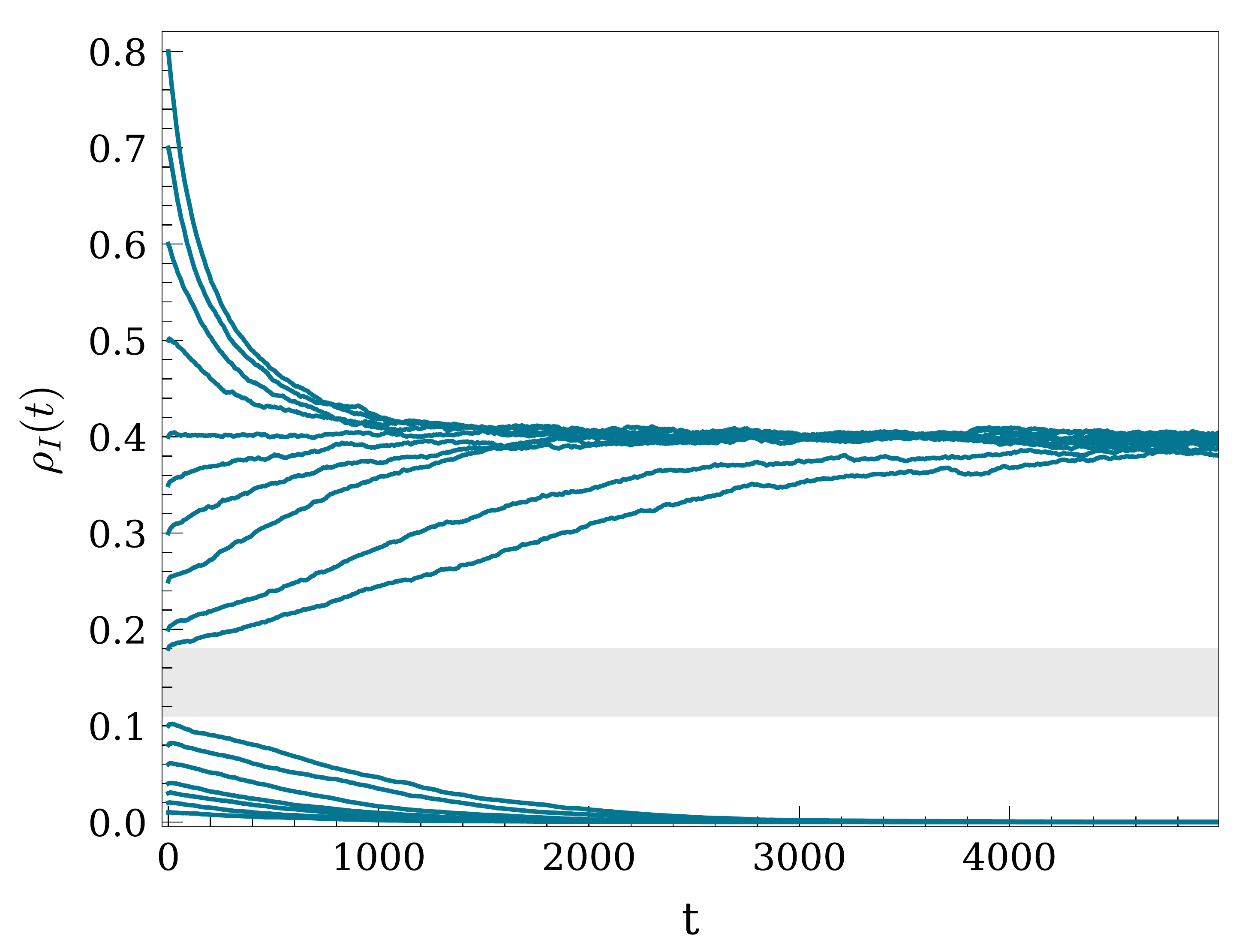}
    \caption{Time evolution of the prevalence, starting from different initial conditions. This shows the bistability. The light gray stripe shows the region at which stochasticity allows the system to go both directions, causing the average curve to lay in between them. Parameters are: $\beta = 0.0009$, $\mu = 0.005$, $r = 1$, $v = 0.3$, $N = 1600$ and $L = 20$ (thus $\khomix = 12.57$). Each curve is an average over $\nexec = 50$ executions.}
    \label{fig:attrac_basins}
\end{figure}

To show that our model presents bistability in the high density regime, we plot, in figure \ref{fig:attrac_basins}, the time series of the \lr{} model for different initial infected fractions $\rho_I(0)$, averaged over $\nexec = 50$ executions each, and observe the basins of attraction. The system can converge to two different stationary states, depending on the initial conditions. For this figure, we use $N = 1600$ individuals in the population to reduce stochastic effects. Yet, the gray-shaded region represents the approximate location of an unstable fixed point where, due to stochastic fluctuations, each execution of the simulation can take different courses.

\begin{figure}
    \centering
    \includegraphics[width=0.95\columnwidth]{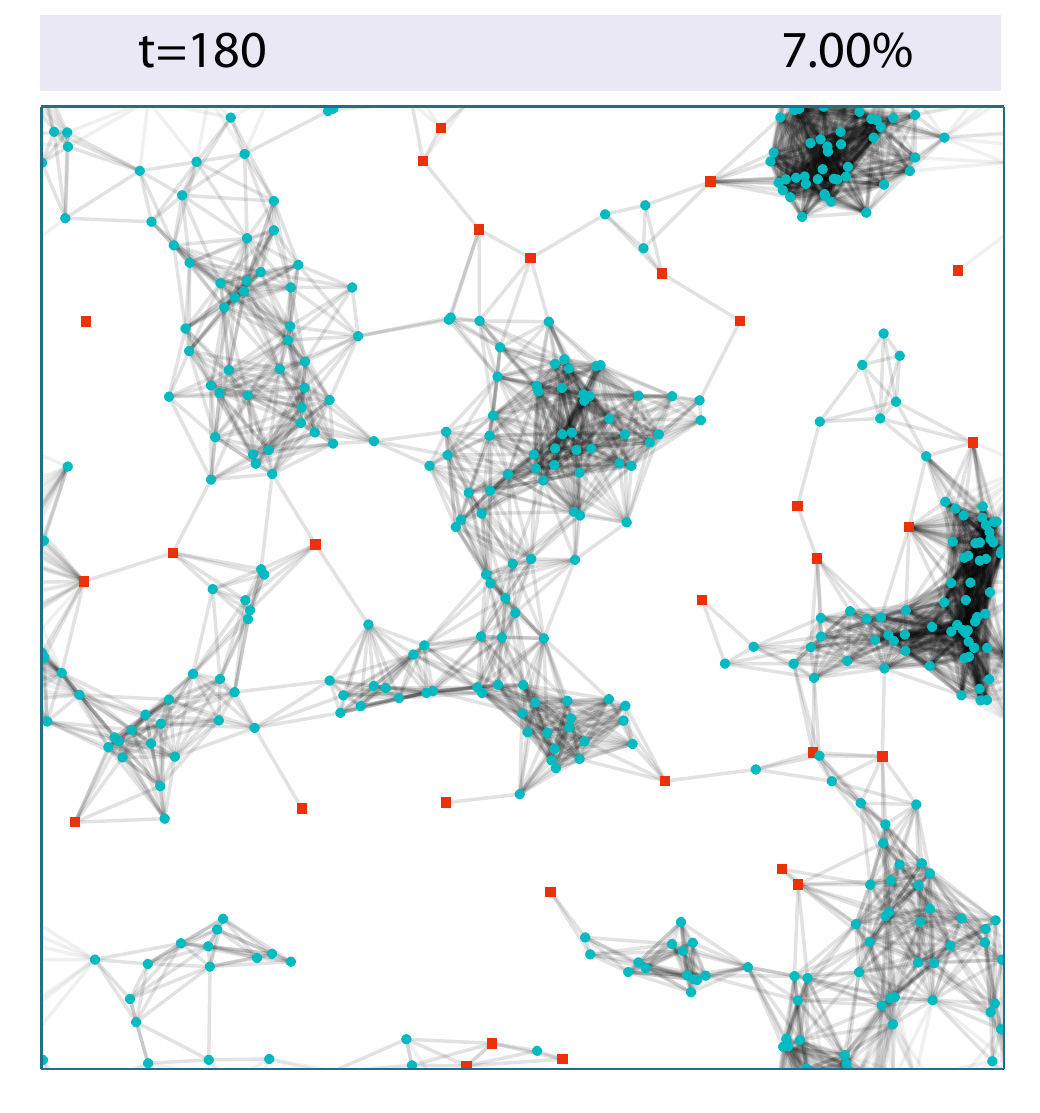}
    \caption{Snapshot network of the population with 400 agents after $t = 180$ steps of a simulation with $\beta = 0.001$, $\mu = 0.005$, $r = 1$, $v = 0.3$ and $\khomix = 12.57$. Red squares represent infected agents, while blue circles are susceptible. The average degree of susceptibles is $\mean{k_S} = 21.5$ and of infecteds is $\mean{k_I} = 9.4$. The prevalence is around $7\%$.}
    \label{fig:flocks_ex}
\end{figure}

\begin{figure}
    \centering
    \includegraphics[width=0.95\columnwidth]{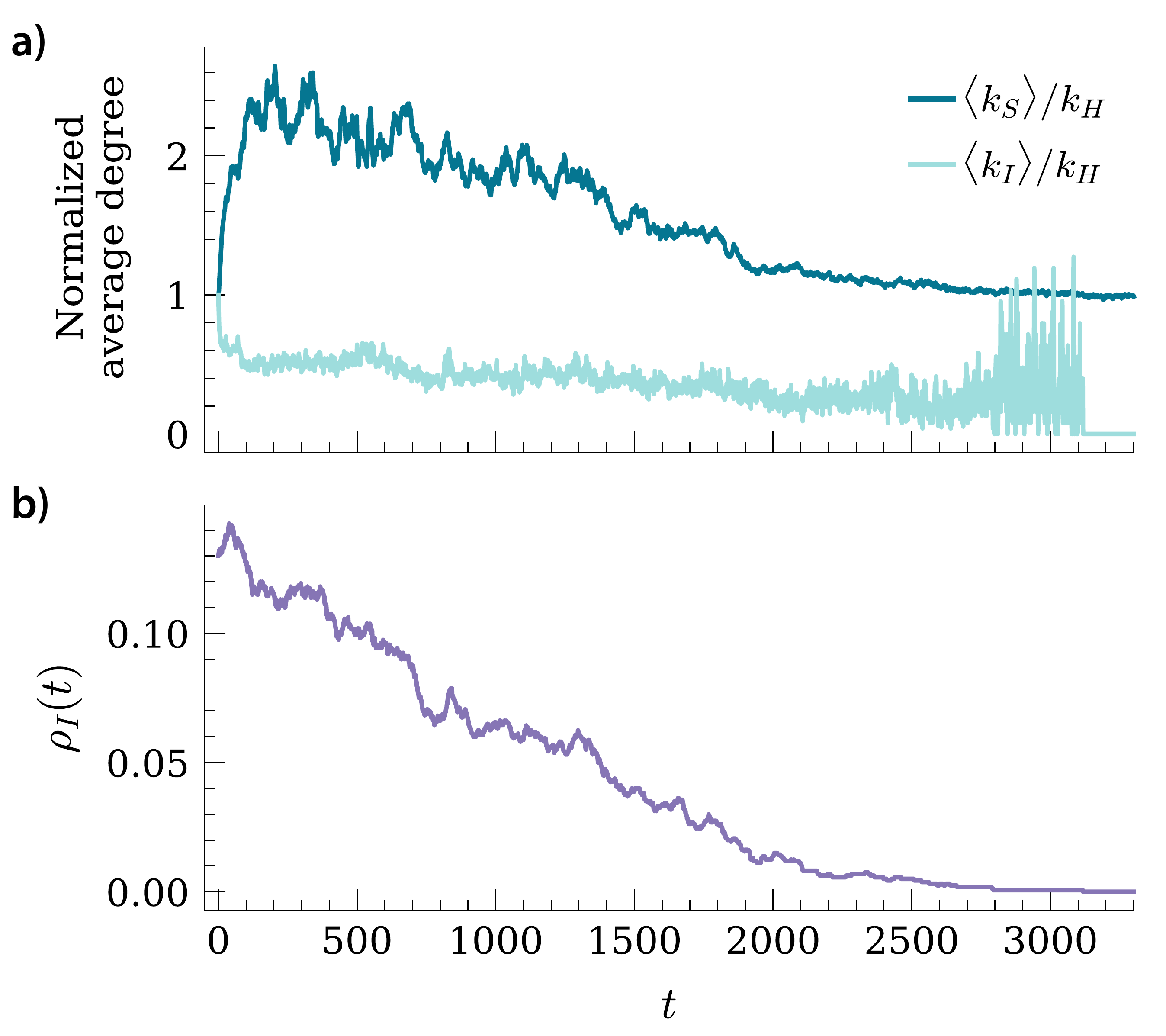}
    \caption{Time evolution of the average degree of susceptible (blue) and infected (orange) agents for a single execution, shown in panel a), along with the prevalence in panel b). The degrees are normalized by the homogeneous degree $\khomix$. Parameters are the same as in figure \ref{fig:attrac_basins}, with $\rho_I(0) = 0.13$ and a single $\nexec = 1$ execution.}
    \label{fig:dyn_metrics}
\end{figure}

As reported in previous works with adaptive networks \cite{gross2006epidemic,risau2009contact,zhou2012epidemic}, susceptible individuals can form highly connected clusters due to the behavioral response mechanism. We report such feature in our model by noticing that, during transient stages of simulations under high density regime, susceptible agents form spatial clusters that are densely connected due to proximity. Figure \ref{fig:flocks_ex} shows a ``snapshot'' network of the population during transient stage of a typical execution, clearly showing the presence of clusters of susceptibles (blue circles), which we call \emph{S-clusters}. The time evolution of this transient behavior can be captured by the average degrees (normalized by $\khomix{}$) among susceptible and infected agents, as shown in figure \ref{fig:dyn_metrics} (a), along with the prevalence (b) for a single execution starting with $\rho_I(0) = 0.13$. As the simulation starts, the degree of susceptible agents (blue curve) quickly raises as the S-clusters are formed, whereas the degree of infected agents drops. For this particular execution, the avoidance mechanism was able to reduce and eventually eliminate the disease, but the initial prevalence falls into the shaded region in figure \ref{fig:attrac_basins}, meaning that the destiny of the system could have been different. 

The formation of S-clusters is related to the spatial gaps that are left by the infected agents performing simple random walks. Susceptible agents tend to move into such gaps and form the observed clusters, which move and change through time.



In other works, clusters of susceptibles were observed both when S individuals actively seek connections to other susceptibles \cite{gross2006epidemic,guerra2010annealed} and when they simply avoid infectious contacts \cite{risau2009contact}. Our model is an example of the latter, reinforcing that the S-clusters can occur without explicit preference to susceptibles in the rewiring mechanism.

The S-clusters are transient. If the prevalence at a given time and the disease transmissibility are not high enough, the lack of infective contacts causes the disease to disappear, along with the S-clusters. However, if the prevalence and/or transmissibility are high enough, the infected agents eventually break the S-clusters and the disease takes over the population, which reaches a steady non-clustered regime. The break of the susceptible clusters is similar to that observed by Zhou and others \cite{zhou2012epidemic} during the epidemic bursts present in their model. 

Due to this ambiguity of trajectories from the clustered regime, it is clear that such clusters are closely related to our model's bistability. In the next section, we develop a semi-analytical approach that corroborates this hypothesis. 



\section{A semi-analytic approach for slow epidemics}

For slow epidemic evolution, achieved by sufficiently low values of $\beta$ and $\mu$, we can not only approximate the discrete time dynamics of the disease to continuous, but also assume that any metrics related to the population of agents is a function of the prevalence, as the population quickly responds to the slow changes in epidemic states. With this in mind, the SIS dynamics is given by the following rate equation for the overall prevalence $\rho_I$:

\begin{equation}
    \label{eq:xdot}
    \dot{\rho_I} = \beta l_{SI}(\rho_I) - \mu \rho_I
\end{equation}
\noindent
where $l_{SI} = L_{SI} / N$ is the number of links $L_{SI}$ that connect susceptible to infected agents normalized by the population size $N$, and is a function of the prevalence $\rho_I$. Effectively, this approximation promotes a time scale decoupling of the epidemic and motion dynamics.

\begin{figure}[!ht]
    \centering
    \includegraphics[width=0.95\columnwidth]{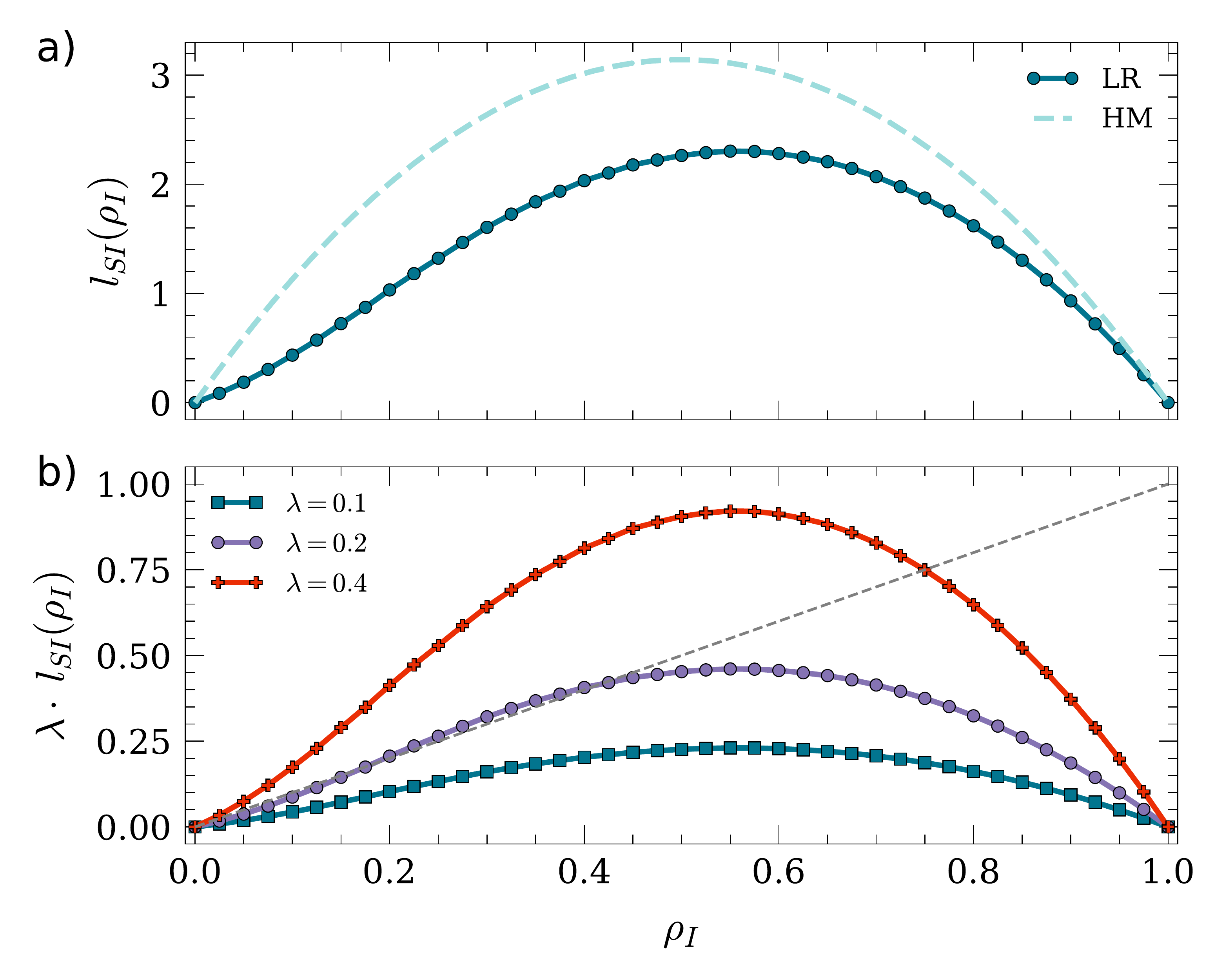}
    
    \caption{Panel a): $l_{SI}$ as a function of the (static) prevalence $\rho_I$, for the local reaction agents (\lr) and for homogeneous mixed populations (\homix). Panel b): $\lambda \cdot l_{SI}$ curve of \lr{} model for three values of the infection-to-healing probability ratio $\lambda$, along with the identity line (gray dashed).  We use $N = 400$, $r = 1$, $v = 0.25$, and $\khomix = 12.57$.}
    \label{fig:lsi_curves}
\end{figure}

We know no analytical method to estimate the functional form of $l_{SI}(\rho_I)$ for mobile agents, but we can directly sample it from Monte Carlo simulations of the population with no epidemic dynamics. That is: each agent receives a given state at the beginning, S or I, and holds it during the whole simulation. This way, we can manually set the number of infected agents $N_I$ to achieve the desired value of the prevalence $\rho_I = N_I / N$. We calculate $l_{SI}$ at each time step in the stationary state, then average its value over time and over independent executions. This process is repeated for different values of $\rho_I$, enough to have a precise shape of the $l_{SI}(\rho_I)$ curve, which is then interpolated to obtain a continuous approximation. For this work, we apply a spline interpolation of third order. Figure \ref{fig:lsi_curves} shows the data acquired by this method. In the a) panel, we see how the \lr{} model deviates from the mass-action law used for homogeneous mixing \cite{fofana2017mechanistic}.

\begin{figure}[!ht]
    \centering
    \includegraphics[width=0.95\columnwidth]{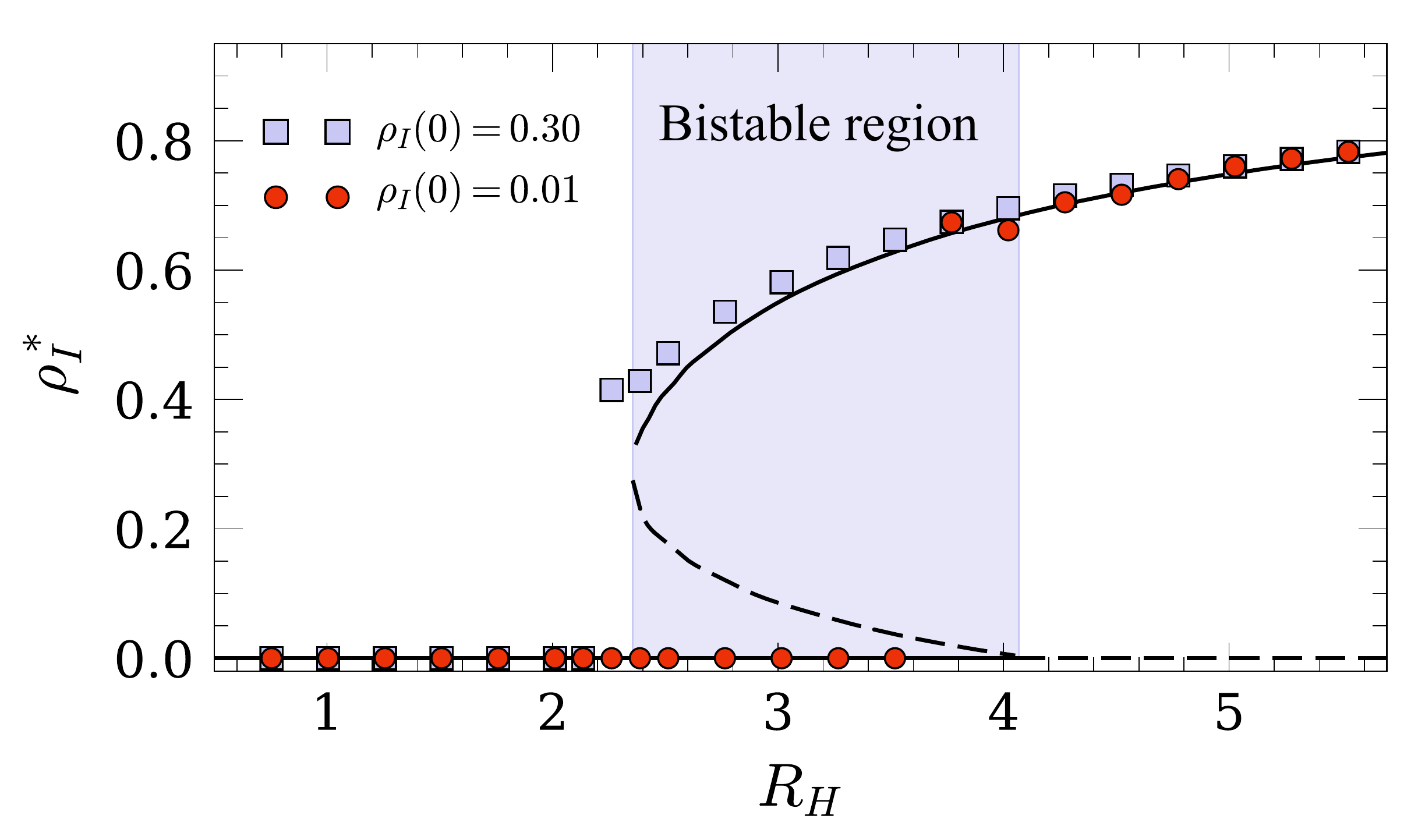}
    \caption{Comparison between the fixed points of $\lambda \cdot l_{SI}$ using the semi-analytical approach (solid lines are stable solutions, dashed ones are unstable) and the results of Monte Carlo simulations, using $\rho_I(0) = 0.3$ (blue squares) and $0.01$ (red circles) as the initial prevalence. We use $N = 400$ for the $l_{SI}$ curves and $N = 1600$ for the Monte Carlo simulations, $r = 1$, $v = 0.25$, $\khomix = 12.57$ and $\mu = 0.005$. Each point is an average over $\nexec = 50$ executions with $\ttransient = 5000$ and $\taverage = 10000$. Different values of $\Rhomix$ are obtained by varying the infection probability $\beta$.}
    \label{fig:prev_comparison}
\end{figure}

Once obtained, the function $l_{SI}(\rho_I)$ can be used to determine the epidemic dynamics in the slow regime. For instance, the fixed points are the values $\rho_I^*$ for which $\dot{\rho_I} = 0$ in equation \ref{eq:xdot}. This is equivalent to solving the equation:

\begin{equation}
    \rho_I^* = \lambda l_{SI}(\rho_I^*)
\end{equation}
\noindent
where $\lambda = \beta / \mu$ is the infection-to-healing rates ratio. 
In figure \ref{fig:lsi_curves}.b), the fixed points can be found as the crossings between $\lambda l_{SI}(\rho_I)$ and the identity line (gray dashed line). Also according to equation \ref{eq:xdot}, the stability of the solution is given by the slope of $\lambda l_{SI}(\rho_I)$ at the fixed point: if it is greater than 1 (i.e., the slope of the identity line), the solution is unstable, and, if it is less than 1, it is stable.

Notice from figure \ref{fig:lsi_curves}.b) that each curve represents a different phase of the \lr{} model. For $\lambda = 0.1$, the only solution is the trivial $\rho_I = 0$, i.e., the healthy state. For $\lambda = 0.2$, the system presents two more solutions, of which one is stable (the one with greater $\rho_I$), while the healthy state remains stable. This characterizes the bistable phase, obtained after a saddle node bifurcation. Finally, for $\lambda = 0.4$, the healthy solution is unstable (as the initial slope of $\lambda l_{SI}(\rho_I)$ is greater than 1), characterizing the regular epidemic phase. This phase is reached after a transcritical bifurcation.

To compare the semi-analytical approach with Monte Carlo simulations of the epidemic dynamics, we plot the fixed points (both stable and unstable) obtained from the $l_{SI}(\rho_I)$ curves as a function of the disease transmission rate $\beta$, rescaled as the homogeneous reproduction number $\Rhomix = \beta \khomix / \mu$. The results are in figure \ref{fig:prev_comparison}, where fixed points of the semi-analytical approach are represented as solid (stable) and dashed (unstable) lines, and the stationary prevalences obtained from Monte Carlo simulations are represented as squares (starting from $\rho_I(0) = 0.30$) and circles (starting from $\rho_I(0) = 0.01$). The region where both the healthy and endemic solutions are stable (according to the semi-analytical approach) is shaded in the plot. 

Figure \ref{fig:prev_comparison} shows the good agreement between the semi-analytical and Monte Carlo formulations. Notice that, in the bistable region, the stationary prevalence of the Monte Carlo simulations depends on the initial conditions, although there is some disagreement at the edges of the bistable region, which may be attributed to stochasticity and population's finite size. 

\begin{figure}[!ht]
    \centering
    \includegraphics[width=0.95\columnwidth]{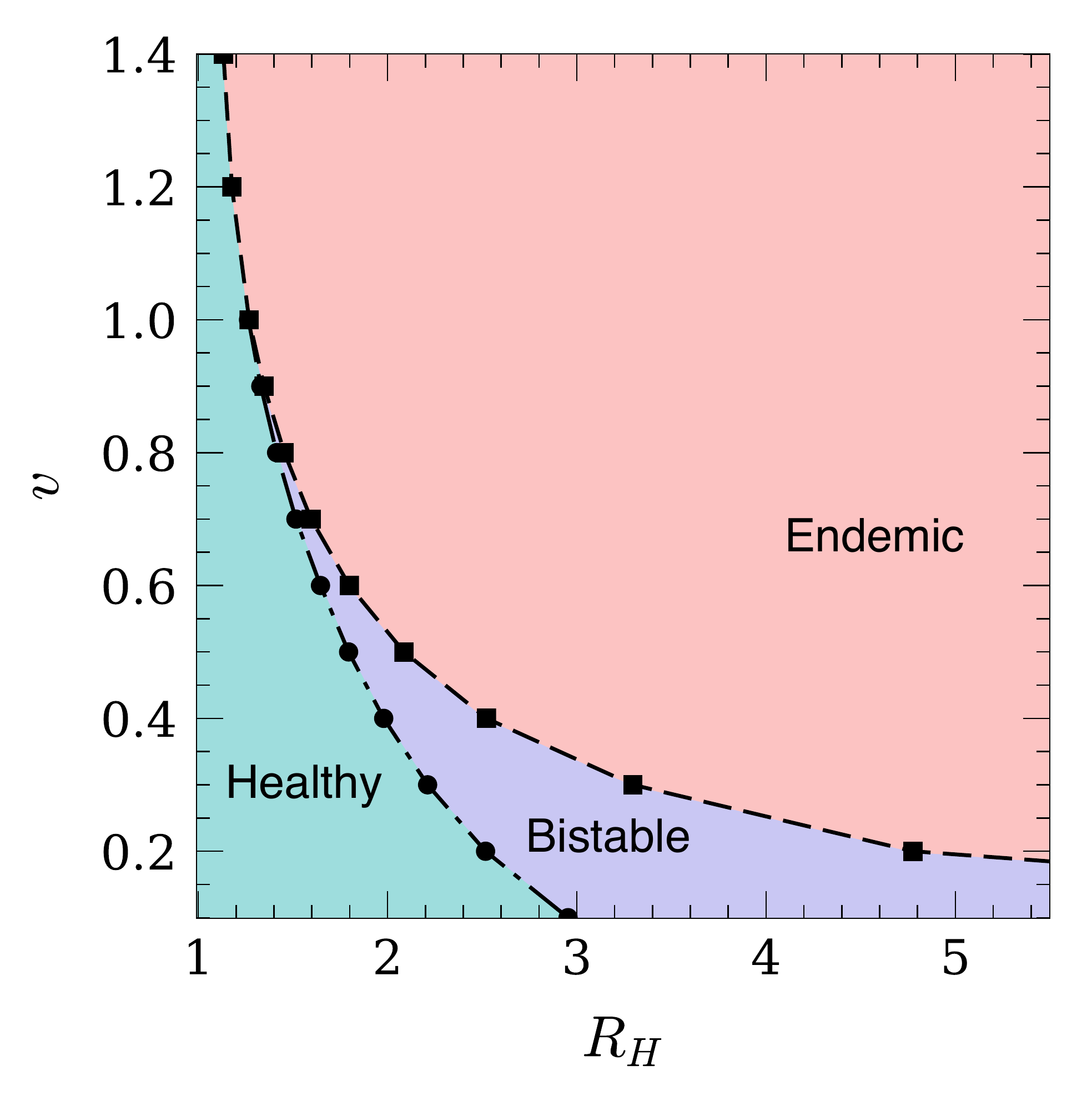}
    \includegraphics[width=0.95\columnwidth]{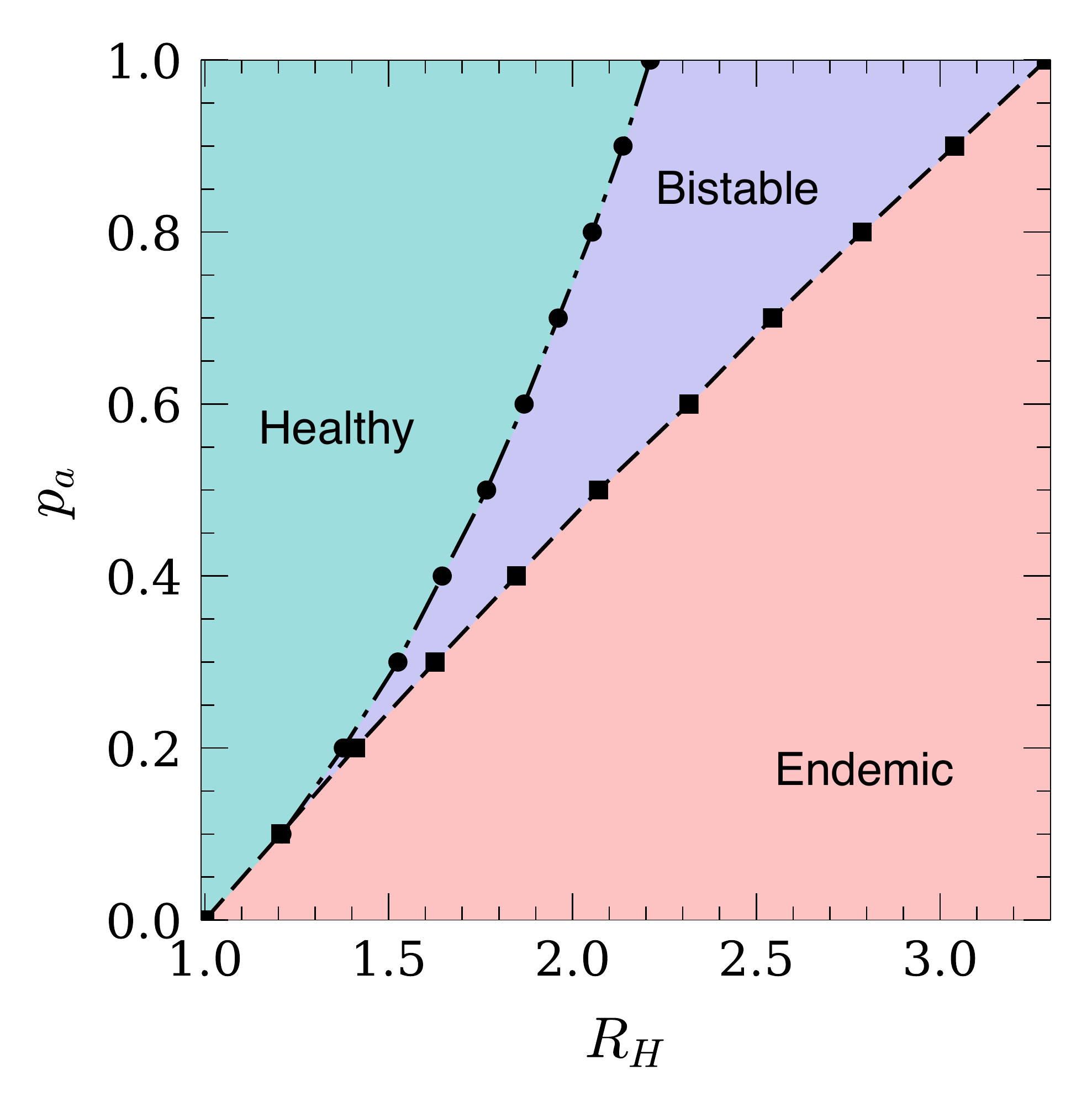}
    \caption{Phase diagrams of the mobile agents model with local reaction (\lr) mechanism, obtained using the semi-analytical approach. The \emph{y} axis parameter is the agent's step size $v$ in the upper panel, and the local reaction probability $\pavoid$ in the lower panel, while different values of $\Rhomix$ are obtained by varying the transmission probability $\beta$. A dashed line with squares represents a transcritical bifurcation, while a dash-dotted line with circles is a saddle-node bifurcation. Other parameters are: $N = 400$, $r = 1$, $v = 0.3$ (lower panel) and $\pavoid = 1$ (upper panel).}
    \label{fig:phase_diagram}
\end{figure}

Using the sampled and interpolated approximation for $l_{SI}(\rho_I)$, we can numerically calculate the critical points. In figure \ref{fig:phase_diagram}, we show two phase diagrams: one for the agents' step size $v$ (upper panel) and another for the local reaction parameter $\pavoid$, both as a function of the homogeneous reproduction number $\Rhomix$.

As the velocity $v$ increases, the bistable region shrinks and disappears, as the fast motion prevents the agents from forming S-clusters. For greater velocities, the critical value of $\Rhomix$ approaches that of a homogeneously mixed population (i.e., $\Rhomix = 1$). From the $\pavoid $ phase diagram (lower panel), we infer that the size of the bistable region grows with the parameter $\pavoid$ that controls the intensity of the local reaction, as expected.

\section{Characterization of the dynamic network}

\begin{figure}[!ht]
    \centering
    \includegraphics[width=0.95\columnwidth]
    {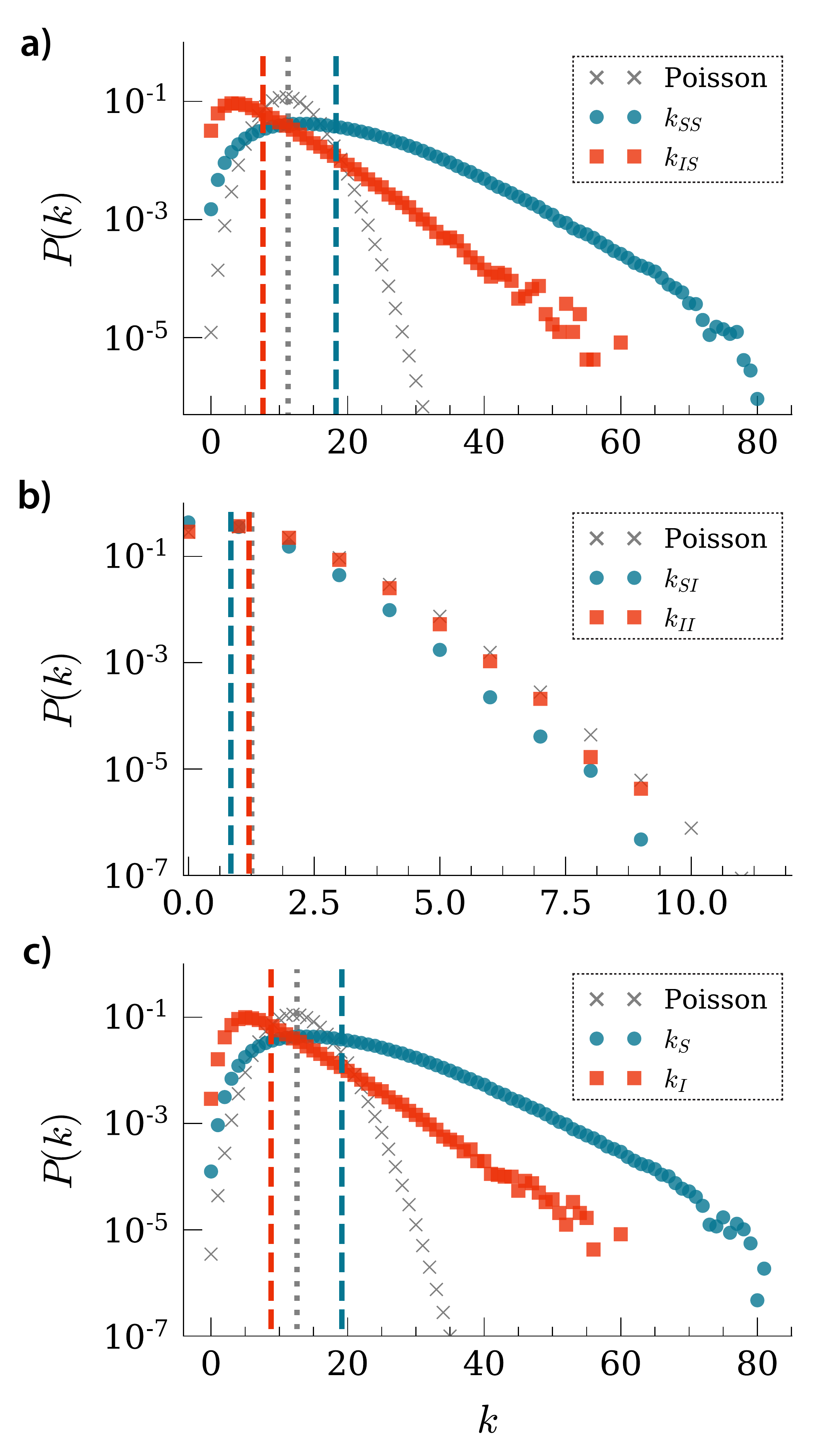}
    \caption{Degree distributions for snapshots of the dynamic network, averaged over time steps and executions, using simulations of the local reaction (\lr) model with static disease states. We group the degrees according to the state of target nodes: susceptibles (a), infecteds (b) and both (c). Each dashed line is the average of the distribution with the same color. As a null model, we show a Poisson distribution (``x'' symbols) that represents homogeneously random interactions in each situation, as explained in text. Other parameters are: $N = 400$, $r = 1$, $v = 0.3$, $\pavoid = 1$, $\khomix = 12.57$ and a fixed prevalence $\rho_I = 0.2$.} 
    \label{fig:degree_distr}
\end{figure}

We can further study the structure of the networks that are formed by this model in the clustered regime by looking at the degree distributions. To simplify the execution, we also remove the epidemic dynamics for this measurement, so the S and I agent states are static. We consider the connectivity between different classes of agents, so for example $k_{SS}$ is the number of links of a susceptible agent that point to other susceptibles, $k_{SI}$ is the number of links of a susceptible that point to infected agents, and so on. Also $k_S$ and $k_I$ are the total degrees of susceptibles and infecteds, respectively. 
Figure \ref{fig:degree_distr} shows each of these degree distributions, grouped by the state of the link targets. As a reference, we also plot a Poisson distribution (gray \emph{x} symbols) $f(k; s) = e^{-s} s^k / {k!}$ with $s = \khomix (1 - \rho_I)$ (a), $s = \khomix \rho_I$ (b) and $s = \khomix$ (c). These are, in each case, the expected degree distributions if the agents were homogeneously distributed at random in the space. Each vertical line also shows the average of each distribution of the corresponding color.

From figure \ref{fig:degree_distr}.a), we can see how susceptibles are highly connected to each other, but weakly connected to infected agents, as the $k_{SS}$ distribution spans over higher values than the $k_{IS}$. In panel b), we see that $k_{II}$ is well described by the equivalent Poisson distribution because infected agents perform simple random walks, whereas $k_{SI}$ is slightly reduced due to the local reaction mechanism. Finally, panel c) shows how the overall degree distributions are distorted and broadened from the basic Poisson curve, similar to the effect observed by Gross and others \cite{gross2006epidemic} in adaptive networks. It also shows that, on average, susceptible agents are more connected than infected ones.

\begin{figure}[!ht]
    \centering
    \includegraphics[width=0.95\columnwidth]{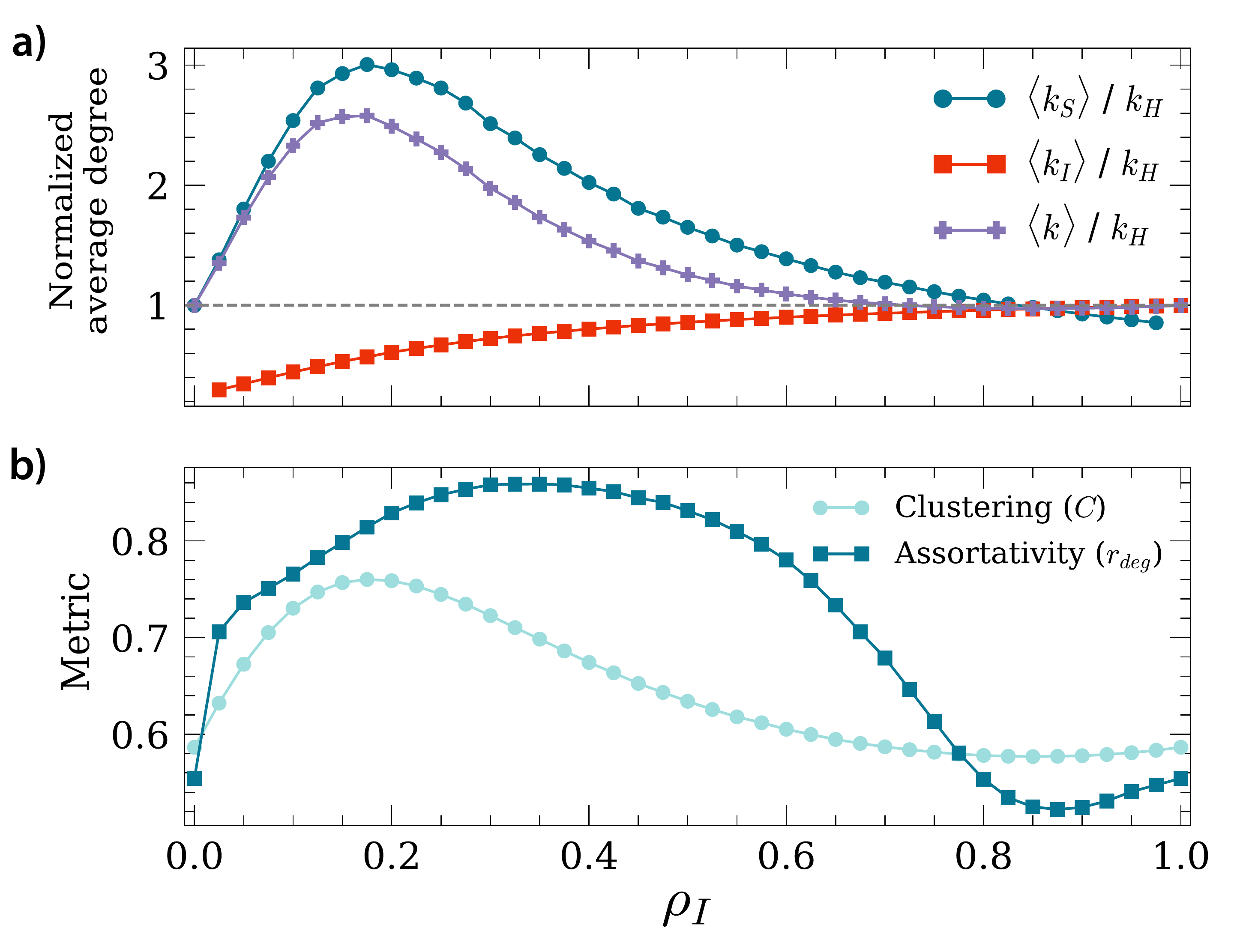}
    \caption{Average degrees normalized by the homogeneous degree $\khomix$ (a) and other metrics (b) as a function of the (static) prevalence $\rho_I$, for snapshot networks of the \lr{} model with no epidemic dynamics ($\beta = \mu = 0$). Other parameters are: $N = 400$, $r = 1$, $v = 0.25$, $\pavoid = 1$, $\khomix = 12.57$. Moreover $\ttransient = 1000$, $\taverage = 10000$, and $\nexec = 40$.}
    \label{fig:metrics_x}
\end{figure}

Yet under the same framework that considers static disease states, we can analyze how the average degrees, as well as some other network metrics, vary with the prevalence, in order to determine where the S-clustering regime occurs. Figure \ref{fig:metrics_x}a shows the average degrees of susceptible ($\mean{k_S}$), infected ($\mean{k_I}$) and ($\mean{k}$) agents of the snapshot network, normalized by the homogeneous degree $\khomix$, as a function of the prevalence. The peak on the $\mean{k_S}$ curve is a consequence of the S-clusters, and it is also visible in the total average degree $\mean{k}$. The average degree of infected agents $\mean{k_I}$, on the other hand, is always smaller than $k_H$, as a consequence of the local reaction mechanism itself, which reduces the contacts with infected agents.

In figure \ref{fig:metrics_x}.b), we show the average clustering coefficient $C$ and the degree assortativity $r_\text{deg}$ of the snapshot networks also as a function of the prevalence. The value of $C$, which is naturally high on dense random geometric networks, is enhanced at the range in which the S-clusters occur, having a good correlation with the $\mean{k_S}$ curve. The assortativity $r_\text{deg}$ is also enhanced by the avoidance mechanism, displaying a broader peak, which means that the effect of the mechanism into the assortativity prevails even when the S-clusters are not very expressive. As also reported in \cite{gross2006epidemic}, the increase in degree correlations may be an effect of the segregation between S and I agents, as reported in the degree distributions in figure \ref{fig:degree_distr}.

We finally notice that the range of unstable prevalences in figure \ref{fig:lsi_curves}.b) is compatible with the region at which $\mean{k_S}$ and $C$ have their peaks in the plots of figure \ref{fig:metrics_x} meaning that, as observed, the S-clusters are unstable when the epidemic dynamics is active, causing the disease to either spread globally by breaking the S-clusters or be eradicated, as explained in section \ref{sec:basic_results}. This reinforces the relationship between the S-clusters and the model's bistability.


\section{Conclusions}

We propose a simple mechanism to include a form of behavioral response to epidemics in mobile agents, based on the avoidance of contacts with infective individuals. We show that, with such mechanism, we can merge the rich dynamical features of adaptive contacts, initially studied on networks, with the overlooked potential of mobile agents for epidemic modeling. Although this work is focused on the high density and slow epidemics regime, we show that different outcomes can be obtained in each regime.

For the low density regime, which is often used to reproduce empirical data \cite{starnini2013modeling}, the local reaction mechanism considerably suppresses the infectious contacts and thus the stationary prevalence, besides increasing the epidemic threshold. This is because, when the agents are spatially sparse, the susceptibles easily find the direction to avoid infected agents. These results, however, might be sensibly affected by considering a ``watch radius'' different from the disease transmission radius, which is proposed as a future work.

In the high density regime, the stationary prevalence is not strongly reduced from the homogeneous mixing scenario, but new dynamical features are introduced: the bistability and the clusters of susceptibles. In the bistable regime, the transient  S-clusters can either succeed to eradicate the disease or permit it to spread, depending on the initial prevalence, disease infectiousness and stochastic factors. The bistability is inherited from the adaptive contact changes, but the collective motion of susceptibles between the gaps left by infecteds is an interesting feature that is exclusive to our spatial model.

We also apply a simple semi-analytic approach to describe the dynamic features of the \lr{} model in the slow epidemics regime, based on simulating the population with static epidemic conditions. Due to spatiotemporal correlations of the random walk network, this approach is less powerful than those proposed on adaptive networks \cite{gross2008robust,marceau2010adaptive,guerra2010annealed}, being unable to capture higher order phenomena such as hysteresis and oscillations that are possibly present in our model too. A more powerful analytical approach can be built by using simulation bursts \cite{gross2008robust,kevrekidis2003equation} to capture higher order moments. Nevertheless, our simple framework can be generalized to a variety of other dynamic population models, provided that the epidemic dynamics is slow, and a functional form of the behavioral reduction of contacts can be extracted from it. 

Finally, we characterize the networks obtained by snapshots of the population with static epidemics in the steady state. With the degree distributions and averages, we show how the local reaction mechanism deviates the behavior from the simple random walk, pointing how the infective contacts are avoided while susceptibles are joint into highly connected clusters. From a practical point of view, this represents a situation in which space is limited and infected individuals do not change their behavior, and is not realistic due to several aspects. However, and as our main goal, we seed the idea of merging adaptive reactions with mobile agent models, calling for further works in the topic.

\section{Acknowledgments}

P.C.V. acknowledges FAPESP for the financial support through grants 2016/24555-0 and 2019/11183-5. Research carried out using the computational resources of the Center for Mathematical Sciences Applied to Industry (CeMEAI) funded by FAPESP (grant 2013/07375-0). F.A.R. acknowledges CNPq (grant 309266/2019- 0) and FAPESP (grant 19/23293-0) for the financial support given for this research. A.A. and Y.M. acknowledge support by Soremartec S.A. Y.M. also acknowledges partial support from the Government of Aragón, Spain through grant E36-20R, and by MCIN/AEI and FEDER funds (grant PID2020-115800GB-I00). The funders had no role in study design, data collection, and analysis, decision to publish, or preparation of the manuscript.

\bibliography{bibliography.bib}

\end{document}